%% file: main.tex
\documentclass[11pt,a4paper]{article}
\usepackage[hyperref]{emnlp2020}
\usepackage{times}
\usepackage{latexsym}
\usepackage{booktabs}
\usepackage{wrapfig}
\usepackage{placeins}

\aclfinalcopy 
\def\aclpaperid{29} 

\usepackage[utf8]{inputenc}

\usepackage{microtype}
\usepackage[compact,small]{titlesec}

\usepackage{times,booktabs,latexsym,amsmath}

\usepackage{url,hyperref}
\usepackage{caption}
\usepackage{subcaption}
\usepackage{graphicx}
\usepackage{balance}  
\usepackage{booktabs}
\usepackage{hyperref}
\usepackage{microtype}
\usepackage{enumitem}
\usepackage{amsmath}
\usepackage{rotating}
\usepackage{relsize}
\usepackage{placeins}

\newif\ifProduction
\Productiontrue

\input{commands.tex}

\title{Flexible retrieval with \nmslib\  and \flexneuart}

\author{
  Leonid Boytsov \Thanks{Work done primarily while at CMU.} \\
  Pittsburgh, PA, USA \\
  \texttt{leo@boytsov.info} \\
  \And
  Eric Nyberg \\
  Carnegie Mellon University\\
  Pittsburgh, PA, USA \\
  \texttt{ehn@cs.cmu.edu}
 }

\date{}

\begin{document}

\maketitle

\begin{abstract}
Our objective is to introduce to the NLP community 
an existing \knn\ search library \nmslib,
 a new retrieval toolkit \flexneuart, 
as well as their integration capabilities.
\nmslib, while being one the fastest \knn\  search libraries,
is quite generic and supports a variety of distance/similarity functions.
Because the library relies on the distance-based structure-agnostic algorithms,
it can be further extended by adding new distances.
\flexneuart\ is a modular, extendible and flexible toolkit for candidate generation
in IR and QA applications, 
which supports mixing of classic and neural ranking signals.
\flexneuart\  can \emph{efficiently} retrieve \emph{mixed} dense and sparse representations 
(with weights learned from training data), which is achieved by extending \nmslib.
In that, other retrieval systems work with purely sparse representations (e.g., Lucene), 
purely dense representations (e.g., \faiss\  and \annoy), or only perform mixing at the re-ranking stage.
\end{abstract}

\input{intro}
\input{nmslib}
\input{flexneuart}

\input{experiments_revised}

\input{conclusion}

\section{Acknowledgements}
This work was done primarily while Leonid Boytsov was a PhD student at CMU
where he was supported by the NSF grant \#1618159.
We thank Sean MacAvaney for making CEDR \cite{macavaney2019cedr} publicly available  
and Igor Brigadir for suggesting to experiment with indexing of BERT word pieces.

\bibliographystyle{acl_natbib}
\bibliography{main}

\balance

\end{document}

%% file: commands.tex
\newcommand{\ttt}{\texttt}
\newcommand{\knnforqa}{\ttt{FlexNeuART}}
\newcommand{\flexneuart}{\ttt{FlexNeuART}}
\newcommand{\faiss}{\ttt{FAISS}}
\newcommand{\annoy}{\ttt{Annoy}}
\newcommand{\onir}{\ttt{OpenNIR}}
\newcommand{\anserini}{\ttt{Anserini}}
\newcommand{\pyserini}{\ttt{Pyserini}}
\newcommand{\terrier}{\ttt{Terrier}}
\newcommand{\pyterrier}{\ttt{PyTerrier}}
\newcommand{\nmslib}{\ttt{NMSLIB}}
\newcommand{\kgraph}{\ttt{KGraph}}
\newcommand{\knn}{$k$-NN}
\newcommand{\tfidf}{TF$\times$IDF}
\newcommand{\manner}{Yahoo Answers Manner}

\usepackage{color}

\ifProduction
\newcommand{\note}[1]{}

\else
\newcommand{\note}[1]{\textbf{\color{red}\small #1}}

\fi

\newcommand{\modelonens}{Model~1}
\newcommand{\modelone}{\modelonens{} }

\usepackage{listings}

\definecolor{lightgray}{rgb}{.9,.9,.9}
\definecolor{darkgray}{rgb}{.4,.4,.4}
\definecolor{forestGreen}{RGB}{34,139,34}
\definecolor{orangeRed}{RGB}{255,69,0}
\lstdefinestyle{workflowStyle}{
language=XML,
alsolanguage=bpel,
alsolanguage=xaml,
basicstyle=\scriptsize,
sensitive=true,
showstringspaces=false,
numbers=left,
numberstyle=\tiny,
tabsize=4,
numbersep=3pt,
extendedchars=true,
xleftmargin=2em,
lineskip=1pt,
breaklines,
captionpos=b,
backgroundcolor=\color{lightgray},
morekeywords={BooleanExpression},
alsoletter={:,,/,?},
morestring=[b]{"},
morecomment=[s]{&lt;!--}{--&gt;},keywordstyle=\color{forestGreen},
identifierstyle=\color{blue}\ttfamily,
stringstyle=\color{orangeRed}\ttfamily,
commentstyle=\color{forestGreen}\ttfamily
}
\lstdefinelanguage{bpel}{
morekeywords={name,linkName,isolated,parallel,partnerLink,operation,portType,inputVariable,createInstance,
variable,element,location,importType,partnerLinkType,myRole,messageType,properties,level,outputVariable,
xmlns,version,encoding}
}
\lstdefinelanguage{xaml}{
morekeywords={TypeArguments,Name,Default,DisplayName,OperationName,ServiceContractName,Key,AddressUri,
CanCreateInstance, LogName, Message, MessageNumber, Expression,CorrelationHandle,Request}
}
\lstdefinelanguage{xml}{
basicstyle=\small,
sensitive=false,
}
\lstnewenvironment{myxml}[2]{
\lstset{style=workflowStyle}
}{}

\usepackage{bera}
\usepackage{listings}
\usepackage{xcolor}

\colorlet{punct}{red!60!black}
\definecolor{background}{HTML}{EEEEEE}
\definecolor{delim}{RGB}{20,105,176}
\colorlet{numb}{magenta!60!black}

\lstdefinelanguage{json}{
    basicstyle=\scriptsize,
    sensitive=true,
    showstringspaces=false,
    numbers=left,
    numberstyle=\tiny,
    tabsize=4,
    numbersep=3pt,
    extendedchars=true,
    xleftmargin=2em,
    lineskip=1pt,
    breaklines,
    captionpos=b,
    backgroundcolor=\color{background},
    literate=
     *{0}{{{\color{numb}0}}}{1}
      {1}{{{\color{numb}1}}}{1}
      {2}{{{\color{numb}2}}}{1}
      {3}{{{\color{numb}3}}}{1}
      {4}{{{\color{numb}4}}}{1}
      {5}{{{\color{numb}5}}}{1}
      {6}{{{\color{numb}6}}}{1}
      {7}{{{\color{numb}7}}}{1}
      {8}{{{\color{numb}8}}}{1}
      {9}{{{\color{numb}9}}}{1}
      {:}{{{\color{punct}{:}}}}{1}
      {,}{{{\color{punct}{,}}}}{1}
      {\{}{{{\color{delim}{\{}}}}{1}
      {\}}{{{\color{delim}{\}}}}}{1}
      {[}{{{\color{delim}{[}}}}{1}
      {]}{{{\color{delim}{]}}}}{1},
}

%% file: intro.tex
\section{Introduction}
Although there has been substantial progress on machine reading tasks using neural models such as BERT \cite{devlin2018bert}, these approaches have practical limitations for open-domain challenges, 
which typically require (1) a retrieval and (2) a re-scoring/re-ranking step to restrict the number of candidate documents.
Otherwise, the application of state-of-the-art machine reading models to large document collections would be impractical
even with recent efficiency improvements \cite{KhattabColBERT2020}.

The first retrieval stage is commonly referred to as the \emph{candidate generation} (i.e., 
we generate candidates for re-scoring).
Until about 2019, the candidate generation would exclusively rely on a traditional search engine such as Lucene,\footnote{\url{https://lucene.apache.org/}}
which indexes occurrences of individual terms, their lemmas or stems \cite{manning2010introduction}.
In that, there are several recent papers where promising results were achieved by generating dense embeddings
and using a \knn\   search library to retrieve them \cite{lee2019latent,karpukhin2020dense,xiong2020approximate}.
However, these studies typically have at least one of the following flaws: 
(1) they compare against a weak baseline such as untuned BM25 or (2) they rely on exact \knn\   search,
thus, totally ignoring practical efficiency-effectiveness and scalability
trade-offs related to using \knn\   search, see, e.g., \S 3.3 \mbox{in~\citet{boytsov2018efficient}}.
\flexneuart\ implements some of the most effective non-neural 
ranking signals: It produced best non-neural runs in the TREC 2019 deep learning challenge \cite{craswell2020overview} 
and would be a good tool to verify these results.

Furthermore, there is evidence that when dense representations perform well, 
even better results may be obtained by combining them with traditional sparse-vector models \cite{seo2019real,gysel2018neural,karpukhin2020dense,kuzi2020leveraging}. 
It is not straightforward to incorporate these representations into existing toolkits,
but \flexneuart\  supports dense and dense-sparse representations
 out of the box with the help of \nmslib\  \cite{DBLP:conf/sisap/BoytsovN13,naidan2015non}.\footnote{\url{https://github.com/nmslib/nmslib}}
\nmslib\  is an efficient library for \knn\  search on CPU, which supports a wide range of similarity functions and 
data formats.
\nmslib\   is a commonly used library\footnote{\url{https://pypistats.org/packages/nmslib}},
which  was recently adopted by Amazon.\footnote{\url{https://amzn.to/3aDCMtC}}
Because \nmslib\  algorithms are largely distance-agnostic,
it is relatively easy to extend the library by adding new distances. 
In what follows we describe \nmslib, \flexneuart, and their integration  in more detail.
The code is publicly available:
\begin{itemize}
    \item \url{https://github.com/oaqa/FlexNeuART}
    \item \url{https://github.com/nmslib/nmslib}
\end{itemize}

%% file: nmslib.tex
\section{\nmslib}\label{SectNMSLIB}
Non-Metric Space Library (\nmslib) is an efficient cross-platform similarity search library 
and a toolkit for evaluation of similarity search methods \cite{DBLP:conf/sisap/BoytsovN13,naidan2015non},
which is the first commonly used library with a principled support for non-metric space searching.\footnote{\url{https://github.com/nmslib/nmslib}}
\nmslib\  is an extendible library, which means that is possible to add new search methods and distance functions. \nmslib\  can be used directly in C++ and Python (via Python bindings). In addition, it is also possible to build a query server, which can be used from Java (or other languages supported by Apache Thrift\footnote{\url{https://thrift.apache.org/}}).

\knn\  search is a conceptually simple procedure that consists in finding 
$k$ data set elements that have highest similarity scores (or, alternatively, smallest distances)
to another element called \emph{query}. 
Despite its formulaic simplicity, \knn\  search is a notoriously difficult problem,
which is hard to do efficiently, i.e., faster than the brute-force scan of the data set,
for high dimensional data and/or non-Euclidean distances.
In particular, 
for some data sets exact search methods do not outperform the brute-force search
in just a dozen of dimensions  (see, e.g., a discussion in \S~1 and \S~2 of Boytsov \citeyear{boytsov2018efficient}).

For sufficiently small data sets and simple similarities, e.g., $L_2$,
the brute-force search can be a feasible solution,
especially when the data set fits into a memory of an AI accelerator.
In particular, the Facebook library for \knn\ search \faiss\ \cite{faiss2017} 
supports the brute-force search on GPU~\footnote{\url{https://github.com/facebookresearch/faiss/wiki/Running-on-GPUs}}.
However, GPU memory is quite limited compared to the main RAM.
For example, the latest A100 GPU has only 40 GB of memory\footnote{https://www.nvidia.com/en-us/data-center/a100/}
while some commodity servers have 1+ TB of main RAM.

In addition, GPUs are designed primarily for dense-vector manipulations 
and have poor support for sparse vectors \cite{hong2018efficient}.
When data is very sparse, as in the case of traditional text indices,
it is possible to efficiently retrieve data using search toolkits such as Lucene.
Yet, for less sparse sets, more complex similarities, 
and large dense-vector data sets
we have to resort to \emph{approximate} \knn\  search,
which does not have accuracy guarantees.

One particular efficient class of \knn\ search methods
relies on the construction of neighborhood graphs for data set points (see
a recent survey by \citeauthor{shimomura2020survey} (\citeyear{shimomura2020survey}) for a thorough description).
Despite initial promising results were published 
nearly 30 years ago \cite{arya1993approximate},
this approach has only recently become popular 
due to good performance
of \nmslib\   and \kgraph\ \cite{dong2011efficient}\footnote{\url{https://github.com/aaalgo/kgraph}}.

Specifically, two successive ANN-Benchmarks challenges \cite{annbench} were won 
first by our efficient implementation of the Navigable Small World (NSW) \cite{malkov2014approximate}
and then by the Hierarchical Navigable Small World (HNSW) contributed to \nmslib\ by Yury Malkov \cite{malkov2018efficient}.
HNSW performance was particularly impressive.

Unlike many other libraries for \knn\  search,
\nmslib\   focuses on retrieval for generic similarities.
The generality is achieved by relying largely on 
\emph{distance-based} methods:
NSW \cite{malkov2014approximate},
HNSW \cite{malkov2018efficient}, 
NAPP \cite{tellez2013succinct,boytsov2016off}, 
and an extension of the VP-tree \cite{boytsov2013learning,Boytsov_Nyberg_2019}.
Distance-based methods can only use values of the mutual data point distances,
but cannot exploit the structure of the data, 
e.g., they have no direct access to vector elements or string characters.
In addition, \nmslib\  has a simple (no compression) 
implementation of a traditional inverted file, 
which can be used to carry out an \emph{exact} maximum-inner product search on sparse vectors.

Graph-based retrieval algorithms have been shown to work efficiently for a variety of 
non-metric and non-symmetric distances \cite{boytsov2019accurate,boytsov2018efficient,naidan2015permutation}.
This flexibility permits adding new distances/similarities with little effort (as
we do not have to change the retrieval algorithms).
However, this needs to be done in C++, which is one limitation.
It is desirable to have an API where C++ code could call Python-implemented distances.
\nmslib\  supports only in-memory indices and with a single exception all indices are static,
which is another (current) limitation of the library.

There is a number of data format and distances---a combination which we call a \emph{space}---supported
by \nmslib. A detailed description can be found online\footnote{\url{https://github.com/nmslib/nmslib/blob/master/manual/spaces.md}}.
Most importantly, the library supports $L_p$ distances with the norm
$\left\| x \right\| _p = \left(  \sum_{i\in I} |x_i|^p \right) ^{1/p}$, the cosine similarity, and the inner
product similarity.
For all of these,
the data can be both fixed-size ``dense'' and variable-size ``sparse`` vectors.
Sparse vectors can have an unlimited number of non-zero elements and their processing is less efficient
compared to dense vectors. On Intel CPUs the processing is speed up using special SIMD operations.
In addition, \nmslib\  supports the Jaccard similarity, the Levenshtein distance (for ASCII strings), 
and the number of (more exotic) divergences (including the KL-divergence).

The library has substantial documentation
and additional information can be found online\footnote{\url{https://github.com/nmslib/nmslib/tree/master/manual}}.

%% file: flexneuart.tex
\section{\flexneuart}
\input{flex_motiv}
\input{flex_desc}

%% file: flex_motiv.tex
\subsection{Motivation}
Flexible classic and NeurAl Retrieval Toolkit, or shortly \flexneuart\  (intended pronunciation flex-noo-art)
is a modular text retrieval toolkit, which incorporates some of the best classic, i.e., traditional,
information retrieval (IR) signals and provides capabilities for integration with recent neural models. 
This toolkit supports all key stages of the retrieval pipeline, 
including indexing, generation of training data, training the models, candidate generation, and re-ranking. 

\flexneuart\   has been under active development for several years and has been used for our own projects,
in particular, to investigate applicability of \knn\ search for text retrieval \cite{boytsov2016off}.
It was also used in recent TREC evaluations \cite{craswell2020overview}
as well as to produce strong runs on the MS MARCO document leaderboard.\footnote{\url{https://microsoft.github.io/msmarco/\#docranking}}
The toolkit is geared towards TREC evaluations: For broader acceptance we would clearly need
to implement Python bindings and experimentation code at the Python level.

\begin{figure}[!t]
\centering
\includegraphics[width=0.49\textwidth]{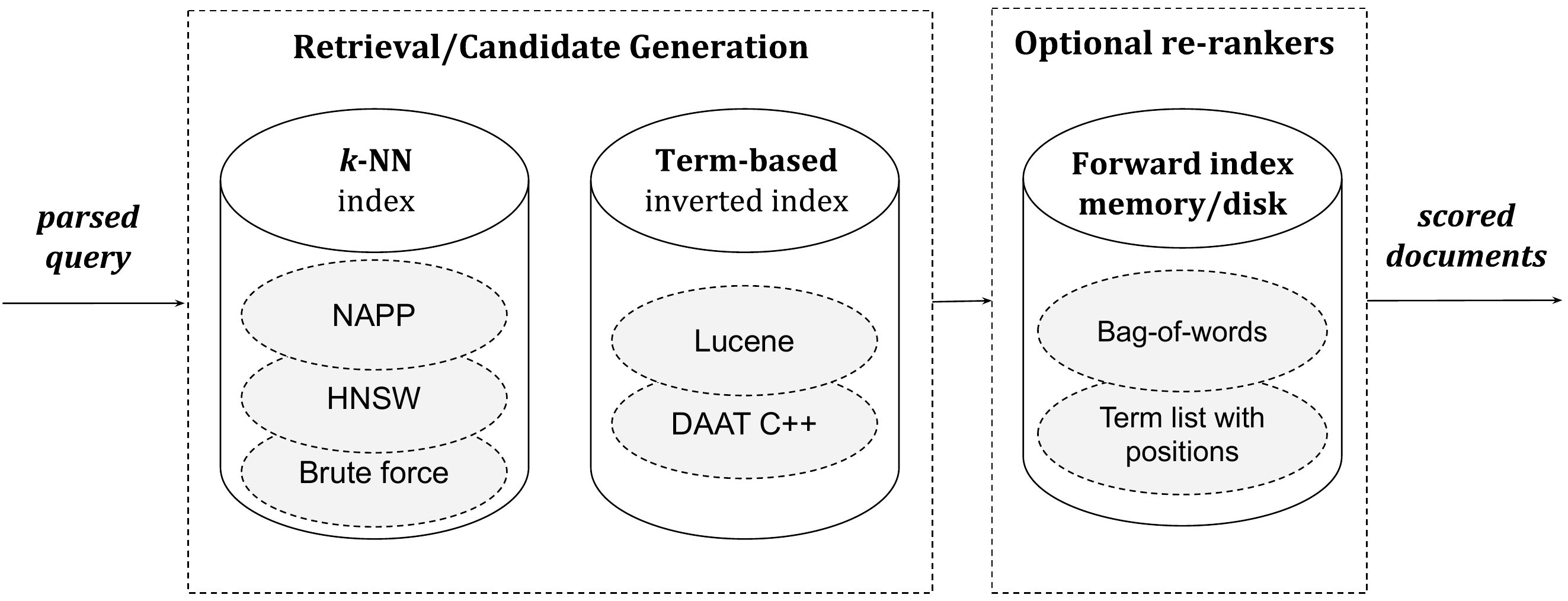}
\caption{Retrieval Architecture and Workflow Overview\label{figureRetrievalDesign}}
\end{figure}

\newpage

\flexneuart\   was created to fulfill the following needs:
\begin{itemize}
    \item \emph{Shallow} integration with Lucene and state-of-the-art toolkits for \knn\  search (i.e.,
          the candidate generation component should be easy to change);
    \item Efficient retrieval and efficient re-ranking with basic relevance signals;
    \item An out-of-the-box support for multi-field document ranking;
    \item An ease of implementation and/or use of most traditional ranking signals;
    \item An out-of-the-box support for learning-to-rank (LETOR) and basic experimentation;
    \item A support for mixed dense-sparse retrieval and/or re-ranking.
\end{itemize}

Packages most similar to ours in retrieval and LETOR capabilities are 
\anserini\  \cite{DBLP:journals/jdiq/YangFL18}, 
\terrier\ \cite{ounis2006terrier}, and \onir\ \cite{macavaney:wsdm2020-onir}. 
\anserini\  and \terrier\  are Java packages, which were recently enhanced with Python bindings
through \pyserini\footnote{\url{https://github.com/castorini/pyserini}} and \pyterrier\ \cite{macdonald2020declarative}.
\onir\  implements re-ranking code on top of \anserini.
These packages are tightly integrated with specific retrieval toolkits, 
which makes implementation of re-ranking components difficult, 
as these components  need to access retrieval engine internals---which are frequently undocumented---to retrieve stored documents, term statistics, etc.
Replacing the core retrieval component becomes problematic as well.
In contrast, our system decouples retrieval and re-ranking modules by keeping an \emph{independent} forward index,
which enables \emph{plugable} LETOR and IR modules.
In addition to this, \onir\  and \pyserini\   do not provide API for fusion of relevance signals 
and none of the toolkits incorporates a lexical translation model \cite{Berger2000},
which can substantially boost accuracy for QA.

%% file: flex_desc.tex
\begin{figure}[t]
\centering
\begin{lstlisting}[language=json]
{
 "DOCNO" : "0",
 "text" : "nfl team represent super bowl 50",
 "text_unlemm" : "nfl teams represented super bowl 50"
}
\end{lstlisting}
\caption{Sample input for question 
``Which NFL team represented the AFC at Super Bowl 50?''\label{figureSampleJSON}
}
\end{figure}

\subsection{System Design and Workflow}\label{sectionDesign}

The \knnforqa\  system---outlined in Figure~\ref{figureRetrievalDesign}---implements a classic multi-stage retrieval pipeline,
where documents flow through a series of ``funnels`` 
that discard unpromising candidates
using increasingly more complex and accurate ranking components.
In that, \flexneuart\  supports one intermediate and one final re-ranker (both are optional).
The initial ranked set of documents is provided by the so-called \emph{candidate generator} (also
known as the \emph{candidate provider}).

\knnforqa\  is designed to work with plugable candidate generators and re-rankers.
Out-of-the-box it supports 
Apache Lucene\footnote{\url{https://lucene.apache.org/}}
and \nmslib, which we describe in~\S~\ref{SectNMSLIB}.
\nmslib\ works as a standalone multi-threaded server implemented with Apache Thrift.\footnote{\url{https://thrift.apache.org/}}
\nmslib\   supports an efficient approximate (and in some cases exact) maximum inner-product search
on sparse and sparse-dense representations.
Sparse-dense retrieval is a recent addition.

Lucene full-text search algorithms rely on classic term-level inverted files, 
which are stored in compressed formats (so Lucene is quite space-efficient).
\nmslib\  (see \S~\ref{SectNMSLIB}) supports
the classic (uncompressed) inverted files with document-at-at-time (DAAT) processing, the brute-force search,  
the graph-based retrieval algorithms HNSW \cite{malkov2018efficient}
and NSW \cite{malkov2014approximate},
as well the pivoting algorithm NAPP \cite{tellez2013succinct,boytsov2016off}. 

The indexing and querying pipelines ingest data (queries and documents) in the form of multi-field JSON entries, 
which are generated by external Java and/or Python code.
Each field can be \emph{parsed} or \emph{raw}.
The parsed field contains \emph{white-space} separated tokens while
the raw field can keep arbitrary text, 
which is tokenized directly by re-ranking components.
In particular, BERT models rely on their own tokenizers \cite{devlin2018bert}.

The core system does not directly incorporate any text processing code,
instead, we assume that an external pipeline does all the processing:
 parsing, tokenization, stopping, and possibly stemming/lemmatization
 to produce a string of white-space separated tokens.
This relieves the indexing code from the need to do complicated parsing
and offers extra flexibility in choosing parsing tools. 

An example of a two-field input JSON
entry for a SQuAD 1.1 \cite{DBLP:conf/emnlp/RajpurkarZLL16}  question
is given in Fig.~\ref{figureSampleJSON}.
Document and query entries contain at least two mandatory fields: \ttt{DOCNO} and \ttt{text},
which represent the document identifier and \emph{indexable} text.
Queries and documents may have additional optional fields. 
For example, HTML documents commonly have a \ttt{title} field.
In Fig.~\ref{figureSampleJSON},
\ttt{text\_unlemm} consists of lower-cased original words, and
\ttt{text} contains word lemmas. 
Stop words are removed from both fields.
From our prior TREC experiments we learned that it is beneficial to 
combine scores obtained for the lemmatized (or stemmed) and the original text \cite{boytsov2011evaluating}.

Retrieval requires a Lucene or an \nmslib\  index, 
each of which can be created \emph{independently}.
To support re-ranking, we also need to create \emph{forward} indices.
There is one forward index for each data field.
For parsed fields,
it contains bag-of-word representations
of documents (term IDs and frequencies) and (optionally) an ordered sequence of words.
For raw fields, the index keeps unmodified text.
A forward index is also required to create an \nmslib\  index.

\begin{figure}[!tb]
\begin{lstlisting}[language=json]
{"extractors": [
  {"type": "TFIDFSimilarity",
   "params": {
    "indexFieldName": "text",
    "queryFieldName": "text",
    "similType": "bm25",
    "k1": "1.2",
    "b": "0.75"}
  },    
  {"type": "avgWordEmbed",
   "params": {
    "indexFieldName": "text_unlemm",
    "queryFieldName": "text_unlemm",
    "queryEmbedFile": "embeds/starspace_unlemm.query",
    "docEmbedFile": "embeds/starspace_unlemm.answer",
    "useIDFWeight": "True",
    "useL2Norm": "True",
    "distType": "l2"}
   }
]}
\end{lstlisting}
\caption{Sample scoring configuration.\label{figureSampleRankJSON}
}
\end{figure}

The \knnforqa\  system has a configurable re-ranking module,
which can combine results from several ranking components.
A sample configuration file shown in Fig.~\ref{figureSampleRankJSON}
contains an array of scoring sub-modules  
whose parameters are specified via nested dictionaries (in curly brackets). 
Each description contains the mandatory parameters \ttt{type} and \ttt{params}.
Scoring modules are feature \emph{extractors}, 
each of which produces one or more numerical feature that can be used by a LETOR component
to train a ranking model or to score a candidate document.

The special composite feature extractor reads the configuration file
and for each description of the extractor it creates an instance of
the feature extractor whose type is defined by \ttt{type}.
The value of \ttt{params} can be arbitrary: parsing and interpreting 
parameters is delegated to the constructor of the extractor object. 

A sample configuration in Fig.~\ref{figureSampleRankJSON}
defines a BM25 \cite{Robertson2004} scorer with parameters $k_1=1.2$
and $b=0.25$ for the index field \ttt{text} (and query field \ttt{text})
as well as the averaged embedding
generator for the fields \ttt{text\_unlemm}.
The latter creates dense query and document representations
using StarSpace embeddings \cite{DBLP:conf/aaai/WuFCABW18}.
There are separate sets of embeddings for queries and documents. 
Word embeddings are weighted using IDFs and subsequently $L_2$ normalized.
Finally, this extractor produces a single feature equal to the
$L_2$ distance between averaged embeddings of the query and the document.

From the forward indices, 
we can export data to NMSLIB and create an index for \knn\  search.
This is supported only for inner-product similarities.
As discussed in the following subsection \S~\ref{sectionDataSimil},
there are two scenarios. 
In the first scenario
we export one vector per feature extractor. 
In particular, we generate a sparse vector for BM25
and a dense vector for the averaged embeddings.
Then, \nmslib\   combines these representations on its own
using adjustable weights, which can be tweaked after data is exported.
In the second scenario--which is more efficient but less flexible---we create one composite vector per document/query,
where individual component weights cannot be changed further after export.

\subsection{Scoring Modules}\label{sectionDataSimil}
Similarity scores between queries and documents
are  computed for a pair of query and a document field
 (typically these are the same fields).\footnote{There can be multiple
 scorers for each pair of fields.}
Scores from various scorers are then combined into a single score by a learning-to-rank (LETOR) algorithm \cite{liu2009learning}.
\knnforqa\ use the LETOR library RankLib from which
we use two particularly effective learning algorithms:
a coordinate ascent \cite{Metzler2007} and LambdaMART \cite{burges2010ranknet}.
We have found a bug in RankLib implementation of
the coordinate ascent: We, thus, use our own, bugfixed, version.

\begin{figure}[!tb]
\begin{lstlisting}[language=json]
[
  {
    "experSubdir": "final_exper",
    "candProvAddConfParam" : "exper_desc/lucene.json",
    "extrType": "exper_desc/final_extr.json", 
    "extrTypeInterm" : "exper_desc/interm_extr.json",
    "modelInterm" : "exper_desc/classic_ir.model",
    "candQty" : 2000,
    "testOnly": 0,
    "runId" : "sample_run_id"
  }
]
\end{lstlisting}
\caption{Sample experimental configuration.\label{figureSampleExperJSON}
}
\end{figure}

Coordinate ascent produces a linear model. It is most effective when the number
of features and/or the number of examples is small.
LambdaMART is a boosted tree model, which, in our experience,
is effective primarily when the number of features and training examples is quite large.

We provide basic experimentation support.
An experiment is described via a JSON descriptor, 
which defines parameters of the candidate generating,
re-ranking, and LETOR algorithms.
Some experimentation parameters such as training and testing subsets
can also be specified in the command line.

A sample descriptor is shown in Fig.~\ref{figureSampleExperJSON}.
It uses an intermediate re-ranker which re-scores
2000 entries with the highest Lucene scores.
A given number of highly scored entries
can be further re-scored using the ``final'' re-ranker.
Note that the experimental descriptor references 
 feature-extractor JSONs rather than
defining everything in a single configuration file.

Given an experimental descriptor,
the training pipeline generates specified features,
 exports results to a special RankLib format and trains the model.
Training of the LETOR model also requires a relevance file (a QREL file in the TREC NIST format),
which lists known relevant documents.
After training, the respective retrieval system is evaluated on another set of queries.
The user can disable model training: This mode is used to tune BM25.

Based on our experience with TREC and community QA collections \cite{boytsov2013learning,boytsov2018efficient},
we support the following scoring approaches:
\begin{itemize}
    \item A proxy scorer that reads scores from one or more standalone scoring servers,
    which can be implemented in Python or any other language supported by 
    Apache Thrift.\footnote{\url{https://thrift.apache.org/}}
    Our system implements neural proxy scorers for CEDR \cite{macavaney2019cedr} and MatchZoo \cite{fan2017matchzoo}.
    We have modified CEDR by providing a better parameterization of the training procedure,
    adding support for BERT large \cite{devlin2018bert} and multi-GPU training.
    \item The \textbf{\tfidf} similarity BM25 \cite{Robertson2004},
    where logarithms of inverse document term frequencies (IDFs) are multiplied by normalized and smoothed
    term counts in a document (TFs).
    \item Sequential dependence model \cite{metzler2005markov}:
    our re-implementation is based on the one from \anserini.
    \item BM25-based proximity scorer, which treats ordered and unordered pairs
    of query terms as a single token. It is similar to the proximity scorer
    used in our prior work \cite{boytsov2011evaluating}.
    \item \textbf{Cosine/$L_2$ distance} between averaged \emph{word} embeddings. 
    We first train word embeddings for the corpus, 
    then construct a dense vector for a document (or query) 
    by applying \tfidf\  weighting to the individual word embeddings and summing them. 
    Then we compare averaged embeddings using the cosine similarity (or $L_2$ distance).
     \item \textbf{IBM Model 1} is a lexical translation model trained using expectation maximization.
     We use Model 1 to compute an alignment log-probability between queries and answer documents. 
     Using Model 1 allows us to reduce the vocabulary gap between queries and documents \cite{Berger2000}.
    \item A proxy query- and document embedder, that produces fixed-size
    dense vectors for queries and documents. The similarity is the 
    inner product between query and document embeddings.
    This scorer operates as an Apache Thrift server.
    \item A BM25-based pseudo-relevance feedback model RM3. Unlike a common approach
    where RM3 is used for query-expansion, we use it in re-ranking mode \cite{diaz2015condensed}.
\end{itemize}

Although \knnforqa\  supports complex scoring models, 
these can be computationally too expensive to be used directly for retrieval  \cite{boytsov2016off,boytsov2018efficient}.
Instead we should stick to a simple vector-space model,
where similarity is computed as the inner product between query 
and document vectors \cite{manning2010introduction}.
The respective retrieval procedure is a maximum inner-product search (a form of \knn\  search).
For example both BM25 and the cosine similarity between
query and document embeddings belong to this class of scorers.

Under the vector-space framework we need to
(1) generate/read a set of field-specific vectors for queries and documents, 
(2) compute field-specific scores using the inner product between
query and document vectors, and (3) aggregate the scores using a linear model.
Alternatively, we can create \emph{composite} queries and document vectors, where we concatenate field-specified vectors multiplied by field weights.
Then, the overall similarity score is computed as the inner product between
composite query and document vectors. 

Our system supports both computation scenarios.
To this end, all inner-product equivalent scorers should inherit from a
specific abstract class and implement
the functions to generate respective query and document vectors.
This abstraction simplifies generation of sparse and sparse-dense query/document vectors, 
which can be subsequently indexed by \nmslib.

%% file: experiments_revised.tex
\input{table_data_sets}

\section{Experiments}\label{SectionExperiments}
We carry out experiments with two objectives:
(1) measuring effectiveness of implemented ranking models;
(2) demonstrating the value of a well-tuned traditional IR system.
We use two recently released MS MARCO collections \cite{craswell2020overview,nguyen2016ms}
and a community question answering (CQA) collection \manner\  \cite{surdeanu2011learning}.
Collection statistics is summarized in Table~\ref{TableDataSets}.

\input{table_small_res}

\input{table_big_res}

MS~MARCO has a document and a passage re-ranking task where 
all queries can be answered using a short text snippet.
There are three sets of queries in each task.
In addition to one large query set with sparse judgments, 
there are two small evaluation sets from the TREC 2019/2020 deep learning track~\cite{craswell2020overview}.
MS MARCO collections query sets were randomly split into training, development  (to tune hyper parameters), 
and test sets.

\manner\ has a large number of paired question-answer pairs.
We include it in our experiments,
because \modelone\   was shown to be effective for CQA data in the past \cite{Jeon2005,RiezlerEtAl2007,surdeanu2011learning,Xue2008}.
It was randomly split into the training and evaluation sets.

Document text is processed using Spacy 2.2.3  \cite{spacy2} to extract tokens and lemmas.
The frequently occurred tokens and lemmas are filtered out using Indri's list of stopwords \cite{strohmanindri2005},
which is expanded to include a few contractions such as ``n't'' and ``'ll''.
Lemmas are indexed using Lucene 7.6.
In the case of MS MARCO documents, entries come in the HTML format.
We extract HTML \ttt{body} and \ttt{title} (and store/index them separately).

In additional to traditional tokenizers,
we also use the BERT tokenizer from the HuggingFace Transformers library \cite{Wolf2019HuggingFacesTS}.
This tokenizer can split a single word into several sub-word pieces \cite{WuSCLNMKCGMKSJL16}.
The stopword list is not applied to BERT tokens.

Training Model 1, which is a translation model, 
requires a parallel corpus where 
queries are paired with respective relevant documents.
The parallel corpus is also known as a \emph{bitext}.
In the case of MS MARCO collections documents are much longer than queries,
which makes it impossible to compute translation probabilities using standard alignment tools \cite{Och2003}.\footnote{\url{https://github.com/moses-smt/mgiza/}}
Hence, for each pair of query $q$ and its relevant document $d$, 
we first split $d$ into multiple short chunks $d_1$, $d_2$, \ldots $d_n$.
Then, we replace the pair $(q, d)$ with a set of pairs $\{ (q, d_i) \}$.

We evaluate performance of several models and their combinations.
Each model name is abbreviated as \ttt{X (Y)}, where \ttt{X}
is a type of the model (see \S \ref{sectionDataSimil} for details) and
\ttt{Y} is a type of the text field. Specifically,
we index original tokens, lemmas, as well as BERT tokens extracted from the 
main document text.
For MS MARCO documents, which come in HTML format, we also extract tokens
and lemmas from the \ttt{title} field.

First, we evaluate performance of the \emph{tuned} \texttt{BM25 (lemmas)}. 
Second, we evaluate fusion models that combine \texttt{BM25 (lemmas)} with 
BM25, proximity, and \modelone\  scores (see \S \ref{sectionDataSimil}) computed 
for various fields.
Note that our fusion models are linear.
Third, we evaluate collection-specific combinations of manually-selected models: 
Except for minor changes these are the fusion models that we used 
in our TREC 2019 and 2020 submissions.

\emph{All models} were trained and/or tuned using training and development sets
listed in Table \ref{TableDataSets}.
For TREC 2019 and 2020 query sets (as well as for \manner), 
the evaluation metric is NDCG@10 \cite{jarvelin2002cumulated},
which the main metric in the TREC deep learning track \cite{craswell2020overview}.
For subsets of MS MARCO collections, we use the mean reciprocal rank (MRR) 
as suggested by Craswell et al.~\citeyearpar{craswell2020overview}.

From the experiments in Table \ref{TableFeatureResults}, 
we can see that for all large query sets the fusion models outperform \texttt{BM25 (lemmas)}.
In particular, the best MS MARCO fusion models are 13-15\% better than
\texttt{BM25 (lemmas)}.
In the case of \manner, 
combining \texttt{BM25 (lemmas)} with Model 1 scores computed for BERT tokens
also boost performance by about 15\%.
For small TREC 2019 and 2020 query sets the gains are marginal.
However,  our fusion models are still better than \texttt{BM25 (lemmas)} by 4-8\%.

We further compare the accuracy of the BERT-based re-ranker \cite{nogueira2019passage}
applied to the output of the tuned traditional IR system with 
the accuracy of the \emph{same} BERT-based re-ranker applied to the output of Lucene (with a BM25 scorer).
The BERT scorer is used to re-rank 150 documents:
Further increasing the number of candidates degraded performance
on the TREC 2019 test set.

By mistake we used the same BM25 parameters for both passages and documents.
As a result, MS MARCO documents candidate generator was suboptimal (passage retrieval
did use the 
properly tuned BM25 scorer). However, we refrained from correcting this error to illustrate
how a good fusion model can produce a strong ranker
via a combination of suboptimal weak rankers.

Indeed, as we can see from Table~\ref{TableCandGen}, 
there is a substantial 4.5-7\% loss in accuracy 
by re-ranking the output of BM25 compared to 
re-ranking the output of the well-tuned
traditional pipeline. 
This degradation occurs in all four experiments.

%% file: table_data_sets.tex
{

\begin{table}[tb]
{
\scriptsize
\begin{tabular}{lccc}\toprule
& \multicolumn{2}{c}{   MS MARCO} &    \begin{tabular}[c]{c} Yahoo \\ Answers \end{tabular} \\
&    documents \hspace{0.25em} &    \hspace{0.25em} passages & \\\midrule
& \multicolumn{3}{c}{   general statistics} \\\midrule
  \# of documents  &     3.2M &     8.8M &     819.6K \\
  \# of doc. lemmas &      476.7  &      30.6  &      20.1  \\
  \# of query lemmas &     3.2  &     3.5  &     11.9  \\ \midrule
& \multicolumn{3}{c}{  \# of queries} \\\midrule
     train/fusion  &     10K  &     20K  &     14.3K  \\
     train/modeling  &     357K  &     788.7K  &     100K  \\
     development  &     2500  &     20K  &     7034  \\
     test  &     2693  &     3000  &     3000  \\
     TREC 2019  &     100  &     100  &       \\
     TREC 2020  &     100  &     100  &       \\
\midrule
& \multicolumn{3}{c}{     BITEXT tokens} \\\midrule
    \# of QA pairs &   43.9M &   4M &   572.8K \\
  \# of query tokens &   2.7 &   2.8 &   12.6  \\
  \# of doc. tokens &   4.3 &   4.2 &   20  \\
\midrule
& \multicolumn{3}{c}{     BITEXT BERT word pieces} \\\midrule
    \# of QA pairs &   50M &   9.5M &   572.8K \\
  \# of query tokens &   6.1 &   2.8 &   42.3  \\
  \# of doc. tokens &   9.4 &   4.3 &   62.7  \\
\bottomrule
\end{tabular}
}
\caption{Data set statistics\label{TableDataSets}}
\end{table}
}

%% file: table_small_res.tex
\begin{table}[tb]
\scriptsize
\centering
\begin{tabular}{@{}l|cc|cc@{}}
\toprule
\begin{tabular}{c} candidate \\ generator\end{tabular}
  & \multicolumn{2}{l|}{\begin{tabular}{c} MS MARCO \\ documents\end{tabular}} & \multicolumn{2}{l}{\begin{tabular}{c} MS MARCO \\ passages\end{tabular}} \\ \midrule
 & \multicolumn{1}{l}{ \begin{tabular}{c} TREC \\ 2019\end{tabular} } & \multicolumn{1}{l|}{develop.} & \multicolumn{1}{l}{\begin{tabular}{c} TREC \\ 2019\end{tabular}} & \multicolumn{1}{l}{develop.} \\
\midrule
BM25         & 0.647 & 0.443 & 0.707 & 0.452 \\
Tuned system & 0.693 & 0.472 & 0.739 & 0.480 \\ \midrule
Gain & 7.08\% & 6.39\% & 4.57\% & 6.08\% \\ \bottomrule
\end{tabular}
\caption{The effect of using a more effective candidate generator (evaluation metric is NDCG@10). BM25 is tuned for MS MARCO passages, but not documents.\label{TableCandGen}}
\end{table}

%% file: table_big_res.tex
\begin{table*}[!tb]
\scriptsize
\centering
\begin{tabular}{@{}l|ccc|ccc|c@{}}
\toprule
 & \multicolumn{3}{c|}{   MS MARCO documents } 
 & \multicolumn{3}{c|}{   MS MARCO passages } 
 & \multicolumn{1}{c}{  \begin{tabular}[c]{@{}c@{}}Yahoo\\ Answers\end{tabular}} \\ \midrule
 & \multicolumn{1}{c}{  test } & \multicolumn{1}{c}{  \begin{tabular}[c]{@{}c@{}}TREC \\ 2019\end{tabular}} & \multicolumn{1}{c|}{  \begin{tabular}[c]{@{}c@{}}TREC \\ 2020\end{tabular}} & \multicolumn{1}{c}{test} & \multicolumn{1}{c}{  \begin{tabular}[c]{@{}c@{}}TREC\\ 2019\end{tabular}} & \multicolumn{1}{c|}{  \begin{tabular}[c]{@{}c@{}}TREC\\ 2020\end{tabular}} & \multicolumn{1}{c}{   test} \\ \midrule
 &    MRR &    NDCG@10 &    NDCG@10 &    MRR &    NDCG@10 &    NDCG@10 &    NDCG@10 \\ \midrule
 &  &  &  &  &  &  &  \\
BM25 (lemmas) & 0.270 & 0.544 & 0.524 & 0.256 & 0.522 & 0.516 & 0.152 \\
BM25 (lemmas)+BM25  (BERT tokens) & 0.283 & 0.528 & 0.537 & 0.270 & 0.518 & 0.525 & 0.159 \\
BM25 (lemmas)+BM25  (tokens) & 0.274 & 0.544 & 0.523 & 0.265 & 0.517 & 0.521 & 0.157 \\
BM25 (lemmas)+BM25  (title tokens) & 0.294 & 0.550 & 0.527 &  &  &  &  \\
BM25 (lemmas)+proximity (lemmas) & 0.282 & 0.559 & 0.524 & 0.257 & 0.538 & 0.523 &  \\
BM25 (lemmas)+proximity (tokens) & 0.284 & 0.560 & 0.531 & 0.265 & 0.534 & 0.524 &  \\
BM25 (lemmas)+Model1 (tokens) & 0.283 & 0.548 & 0.535 & 0.274 & 0.522 & 0.567 & 0.160 \\
BM25 (lemmas)+Model1 (BERT tokens) & 0.284 & 0.557 & 0.525 & 0.271 & 0.517 & 0.509 & 0.175 \\
best combination & 0.310 & 0.565 & 0.542 & 0.268 & 0.558 & 0.560 &  \\
\bottomrule
\end{tabular}
\caption{Evaluation of various fusion models.\label{TableFeatureResults}}
\end{table*}

%% file: conclusion.tex
\section{Conclusion and Future Work}
We present to the NLP community 
an existing \knn\ search library \nmslib,
a new retrieval toolkit \flexneuart, 
as well as their integration capabilities,
which enable efficient retrieval of sparse and sparse-dense document representations.
\flexneuart\  implements a variety of effective traditional relevance signals,
which we plan to use for a fairer comparison with recent neural retrieval systems
based on representing queries and documents via fixed-size dense vectors.
